\documentclass[a4paper,11pt]{article}
\usepackage{pos}
\usepackage{aas_macros}
\usepackage{xspace}
\usepackage{multicol}

\newcommand{\mrc}{MRC~0910-208\xspace}

\newcommand{\rxsj}{1RXS~J195815.6-301119\xspace}
\newcommand{\hess}{H.E.S.S.\xspace}
\newcommand{\fermi}{\textit{Fermi}\xspace}
\newcommand{\swift}{\textit{Swift}\xspace}
\newcommand{\nustar}{\textit{NuSTAR}\xspace}

\newcommand{\zmrc}{0.19802}

\newcommand{\zrxsj}{0.119}
\newcommand{\ltmrc}{17.2\xspace}

\newcommand{\ltrxsj}{7.3\xspace}
\newcommand{\sigmrc}{7.0\xspace}

\newcommand{\sigrxsj}{8.8\xspace}
\newcommand{\Ethrmrc}{0.16\xspace}

\newcommand{\Ethrrxsj}{0.18\xspace}
\newcommand{\plmrc}{3.63\xspace}

\newcommand{\plrxsj}{2.78\xspace}
\newcommand{\dplmrc}{0.38\xspace}

\newcommand{\dplrxsj}{0.26\xspace}
\newcommand{\dsysplmrc}{0.38\xspace}

\newcommand{\dsysplrxsj}{0.14\xspace}
\newcommand{\Nmrc}{5.7\xspace}

\newcommand{\Nrxsj}{4.4\xspace}
\newcommand{\dNmrc}{1.0\xspace}

\newcommand{\dNrxsj}{0.6\xspace}
\newcommand{\Neblmrc}{12.6\xspace}

\newcommand{\Neblrxsj}{8.3}
\newcommand{\dNeblmrc}{2.2}

\newcommand{\dNeblrxsj}{1.2}

\newcommand{\pleblmrc}{2.35\xspace}
\newcommand{\pleblrxsj}{2.00\xspace}

\newcommand{\dpleblmrc}{0.50\xspace}
\newcommand{\dpleblrxsj}{0.27\xspace}
\newcommand{\Emrc}{0.36\xspace}

\newcommand{\Erxsj}{0.47\xspace}

\title{Detection of new Extreme BL Lac objects with H.E.S.S. and \textit{Swift} XRT}
 \ShortTitle{Extreme BL Lacs with H.E.S.S. and \textit{Swift}}

\author[a]{Mathieu de Bony de Lavergne}
\author[b]{Tomas Bylund}
\author[c]{Manuel Meyer}
\author*[d]{Angel Priyana Noel}
\author[a]{David A. Sanchez}

\affiliation[a]{Laboratoire d'Annecy de Physique des Particules, Univ. Grenoble Alpes, Univ. Savoie Mont Blanc, CNRS, LAPP, 74000 Annecy, France}
\affiliation[b]{Linnaeus University, Department of Physics and Electrical Engineering,\\
   35195 V\"axj\"o, Sweden}
\affiliation[c]{Institute for Experimental Physics, University of Hamburg,\\
  Luruper Chaussee 149, 22761 Hamburg, Germany}
\affiliation[d]{Obserwatorium Astronomiczne, Uniwersytet Jagielloński,\\
  Orla 171, 30-244 Krakow, Poland}

\forColl{H.E.S.S. and \emph{Fermi}-LAT} 

\emailAdd{lavergne@lapp.in2p3.fr}
\emailAdd{tomas.bylund@lnu.se}
\emailAdd{manuel.meyer@desy.de}
\emailAdd{anoel@oa.uj.edu.pl}
\emailAdd{david.sanchez@lapp.in2p3.fr}

\abstract{Extreme high synchrotron peaked blazars (EHBLs) are amongst the most powerful accelerators found in nature. Usually the synchrotron peak frequency of an EHBL is above $10^{17}\,$Hz, i.e., lies in the range of medium to hard X-rays making them ideal sources to study particle acceleration and radiative processes. EHBL objects are commonly observed at energies beyond several TeV, making
them also powerful probes of gamma-ray absorption in the intergalactic medium. During the last decade, several attempts have been made to increase the number of EHBL detected at TeV energies and probe their spectral characteristics. Here we report new detections of EHBLs in the TeV energy regime, each at a redshift of less than 0.2, by the High Energy Stereoscopic System (H.E.S.S.). Also, we report on X-ray observations of these EHBLs candidates with \textit{Swift} XRT. In conjunction with the very high energy observations, this allows us to probe the radiation mechanisms and the underlying particle acceleration processes.}

\FullConference{37$^{\rm{th}}$ International Cosmic Ray Conference (ICRC 2021)\\
		July 12th -- 23rd, 2021\\
		Online -- Berlin, Germany}

\begin{document}
\maketitle

\section{Introduction}
Blazars are active galactic nuclei (AGNs) that produce jets of relativistic charged particles, which are closely aligned to the line of sight to the observer. 
Their multi-wavelength spectral energy distributions (SEDs) reveal two characteristic bumps; the low-energy bump is attributed to synchrotron emission of electrons and positrons, whereas the production mechanism for the high-energy emission, which can reach TeV energies, is still debated~\cite[see, e.g.,][]{2016ARA&A..54..725M}. It could either be due to inverse Compton scattering or hadron-initiated processes.
In particular, BL Lac objects -- a sub-class of blazars with only weak emission lines in the optical -- with a peak synchrotron frequency above $10^{15}\,\mathrm{Hz}$ (so-called high-frequency-peaked BL Lac objects; HBLs) have been commonly observed at TeV energies with imaging air Cherenkov telescopes.  
Of particular interest are extreme HBLs (EHBLs) with synchrotron peaks above $\approx 10^{17}\,\mathrm{Hz}$~\cite[see,][for a recent review]{2020NatAs...4..124B}.
Such EHBLs have their high-energy peak beyond 1\,TeV and exhibit hard power-law spectra with spectral indices $\Gamma < 2$ both in the soft X-ray and  $\gamma$-ray band~\cite[see, e.g.,][]{2019ApJ...882L...3P}. 
As $\gamma$-ray emission through the self-synchrotron-Compton (SSC) process becomes suppressed at multi-TeV energies due to the Klein-Nishina effect, these sources are promising candidates for hadron-initiated $\gamma$-ray emission scenarios.  
Furthermore, their multi-TeV emission makes them excellent candidates to study the absorption of $\gamma$~rays on the extragalactic background light (EBL) and to search for $\gamma$-ray cascades, which in turn can be used to constrain the intergalactic magnetic field. 

Here, the detection at very-high $\gamma$-ray energies of two EHBLs, namely \mrc (redshift $z = \zmrc$)  and \rxsj  ($z=\zrxsj$), with the High Energy Stereoscopic System (\hess) is reported.
Archival X-ray observations are used to determine the peak of their synchrotron emission. 

\section{\hess observations}

\hess is an array of 5 imaging air Cherenkov telescopes located in the Khomas highlands in Namibia. The array is composed of four telescopes with a mirror diameter of 12 meters called CT1 to CT4, and one with a mirror diameter of 28 meters called CT5.
\hess can detect very-high energy $\gamma$~rays in the energy range between $\approx$50\,GeV up to $\approx100\,$TeV.

The sources considered here were proposed for observations due to either their high synchrotron peak frequency of $\gtrsim 10^{16}\,$Hz or their hard $\gamma$-ray spectrum measured with the \textit{Fermi} Large Area Telescope (LAT) with spectral indices $\Gamma < 2$ as reported in the 4th AGN \fermi  catalog~\cite[4LAC,][]{2020ApJ...892..105A}.
These properties are common features of blazars already detected at TeV energies.\footnote{See, e.g., \url{http://tevcat.uchicago.edu/}} ²
Both \mrc and \rxsj were observed in 2018 in May and September, respectively, resulting in an acceptance corrected lifetime of \ltmrc\,hours and \ltrxsj\,hours. The observations were carried out in wobble mode, with the ON region offset 0.5 ${}^{\circ}$ from the centre of the camera, and had an average zenith angle of 24${}^{\circ}$ for \mrc and 17 ${}^{\circ}$ for \rxsj.

Events detected with at least three of the small-sized telescopes are considered in the data analysis. The energies are reconstructed using the \texttt{ImPACT} method and standard selection cuts~\cite{2014APh....56...26P}, considering only the small-sized telescopes. This lead to the detection of \mrc and \rxsj with 7.0 $\sigma$ and 8.8 $\sigma$ respectively. 
A combined analysis also including CT5 will be presented elsewhere. 
The data analysis has been carried out with \texttt{gammapy} version 0.18.2~\cite{2015ICRC...34..789D}.\footnote{\url{https://docs.gammapy.org}}

\begin{table}[htb]
\centering
    \begin{scriptsize}
    \begin{tabular}{l|cc}
    \hline
    \hline
    {} & \mrc & \rxsj \\
    \hline
    {} & \multicolumn{2}{c}{\hess Results} \\
    \hline
    $T$ (hours) & \ltmrc & \ltrxsj \\
    $S$ ($\sigma$)& \sigmrc & \sigrxsj \\
    $N$ ($10^{-11}\,\mathrm{TeV}^{-1}\,\mathrm{cm}^{-2}\,\mathrm{s}^{-1}$) & $\Nmrc\pm\dNmrc$  & $\Nrxsj\pm\dNrxsj$ \\
    $\Gamma$ & $\plmrc\pm\dplmrc_\mathrm{stat.}\pm\dsysplmrc_\mathrm{sys.}$  & $\plrxsj\pm\dplrxsj_\mathrm{stat.}\pm\dsysplrxsj_\mathrm{sys.}$ \\
    $N_\mathrm{int}$ ($10^{-11}\,\mathrm{TeV}^{-1}\,\mathrm{cm}^{-2}\,\mathrm{s}^{-1}$) & $\Neblmrc\pm\dNeblmrc$  & $\Neblrxsj\pm\dNeblrxsj$ \\
    $\Gamma_\mathrm{int}$ & $\pleblmrc\pm\dpleblmrc_\mathrm{stat.}\pm\dsysplmrc_\mathrm{sys.}$ &  $\pleblrxsj\pm\dpleblrxsj_\mathrm{stat.}\pm\dsysplrxsj_\mathrm{sys.}$ \\
    $E_0$ (TeV) & \Emrc & \Erxsj \\
    $E_\mathrm{thr}$ (TeV) & \Ethrmrc & \Ethrrxsj \\
    \hline
    {} & \multicolumn{2}{c}{\fermi-LAT Results} \\
    \hline
    $N$ ($10^{-12}\,\mathrm{MeV}^{-1}\,\mathrm{cm}^{-2}\,\mathrm{s}^{-1}$) & $9.241\pm0.558$ & $6.114\pm0.417$  \\
    $\Gamma$ & $1.871\pm0.043$ & $1.835 \pm0.046$ \\ 
    $E_0$ (GeV) & 3.383 &  4. 051  \\
    \hline
    {} & \multicolumn{2}{c}{\swift-XRT Results} \\
    \hline
    $T$ (ks) & 6.4 &  11.6 \\
    $N_\mathrm{PL} (10^{-3}\,\mathrm{keV}^{-1}\mathrm{cm}^{-2}\,\mathrm{s}^{-1})$ & $2.55 \pm 0.19$  & $2.59 \pm 0.11$ \\
    $\Gamma$ & $2.26 \pm 0.134$ & $1.96 +/- 0.063$ \\
    $C_\mathrm{PL}$ & 106.53 & 326.89 \\
    $\mathrm{dof}_\mathrm{PL}$ & 134  & 268\\
    $N_\mathrm{LP} (10^{-3}\,\mathrm{keV}^{-1}\mathrm{cm}^{-2}\,\mathrm{s}^{-1})$ & $2.90 \pm 0.25$ & $2.76 \pm 0.13$ \\
    $\alpha$ & $2.07 \pm 0.16$ & $1.79 \pm 0.086$ \\ 
    $\beta$ & $0.97 \pm 0.39$  & $0.53 \pm 0.18$ \\ 
    $C_\mathrm{LP}$ & 100.15 & 318.01 \\
    $\mathrm{dof}_\mathrm{LP}$ & 133 & 267 \\
    $E_\mathrm{peak}$ (keV) & $2.91\pm1.55$& $5.41\pm1.78$  \\
    $\Delta C = C_\mathrm{PL} - C_\mathrm{LP}$ & 6.38 & 8.88\\
    \hline
    {} & \multicolumn{2}{c}{NuSTAR results} \\
    \hline
    $T$ (ks) & -- & 52.8 \\
    $N_\mathrm{PL} (10^{-3}\,\mathrm{keV}^{-1}\mathrm{cm}^{-2}\,\mathrm{s}^{-1})$ & --  & $4.82 \pm 0.30$  \\
    $\Gamma$ & -- & $2.35 \pm 0.03$ \\
    $C_\mathrm{PL}$ & -- & $756.82$ \\
    $\mathrm{dof}_\mathrm{PL}$ & -- & 818 \\
    \hline
    {} & \multicolumn{2}{c}{SSC Modeling Results} \\
    \hline
    $\alpha_1$  &2.5 & 2.5\\ 
   $\alpha_2$ &3.5 & 3.3\\ 
$\log_{10}(\gamma_b)$ & 5.2  & 5.6\\ 
$\log_{10}(\gamma_\mathrm{max})$ & 6.1 & 6.7  \\ 
$N_e (10^{55})$ &3.06 & 7.71 \\ 
$B$ (G) & 0.01 & 0.01 \\ 
$\log_{10}(R / \mathrm{cm})$ & 17.0& 17.13\\ 
$\Gamma_\mathrm{L}$ & 30 &  20\\ 
    \hline
    \end{tabular}
    \end{scriptsize}
    \caption{
    Observation and modeling results. \emph{The first part} of the table summarizes the results obtained with \hess. The acceptance corrected lifetime is denoted with $T$, $S$ is the detection significance (derived with Eq.~(17) of \cite{1983ApJ...272..317L}) above the threshold energy $E_\mathrm{thr}$. $N$ and $N_\mathrm{int}$ ($\Gamma$ and $\Gamma_\mathrm{int}$) are the normalizations (power-law indices) for the observed and absorption corrected spectra, respectively, and $E_0$ is the de-correlation energy. The power-law parameters are derived through a fit to the \hess data between 0.2 and 2\,TeV.
    \emph{The second part} provides the best-fit power-law parameters derived from LAT observations. 
    \emph{The third part} summarizes the \swift XRT results and fit parameters for an absorption corrected power law and log parabola. The c-stat values $C$ and the degrees of freedom (dof) for each x-ray fit are also reported.
    For the X-ray results only, $E_0$ is fixed to 1\,keV.
    \emph{The fourth part} gives the results for an absorption-corrected power law derived from \nustar observations of \rxsj.
    The \emph{bottom part} summarizes the best-fit parameters for the SED modeling, see Sec.~\ref{sec:ssc}. For both sources, $\gamma_\mathrm{min}=1$.
    The integrated electron distribution is given by $N_e$.
    }
    \label{tab:results}
\end{table}

\begin{figure}[htb]
    \centering
    \includegraphics[width=.49\linewidth]{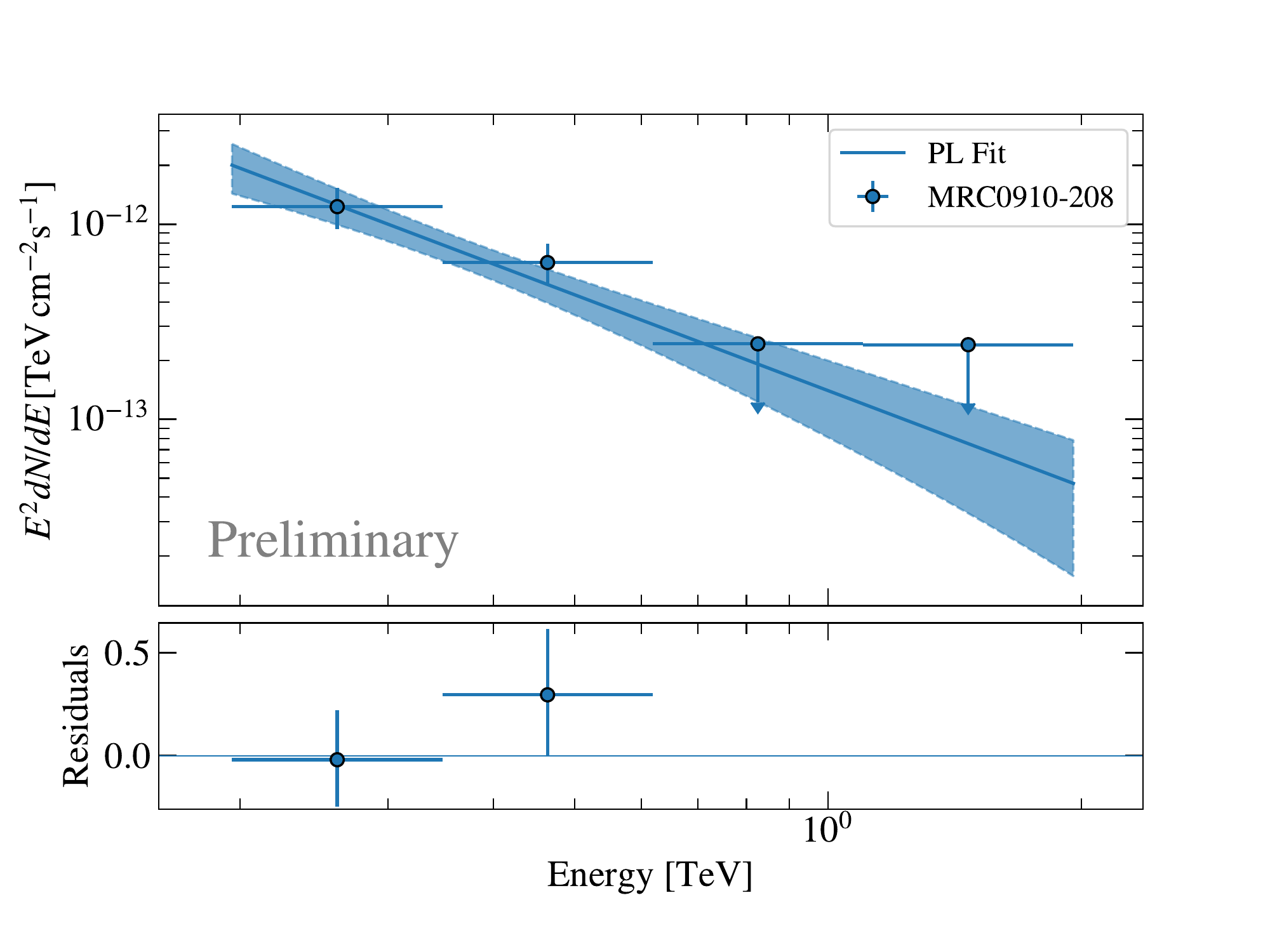}
    \includegraphics[width=.49\linewidth]{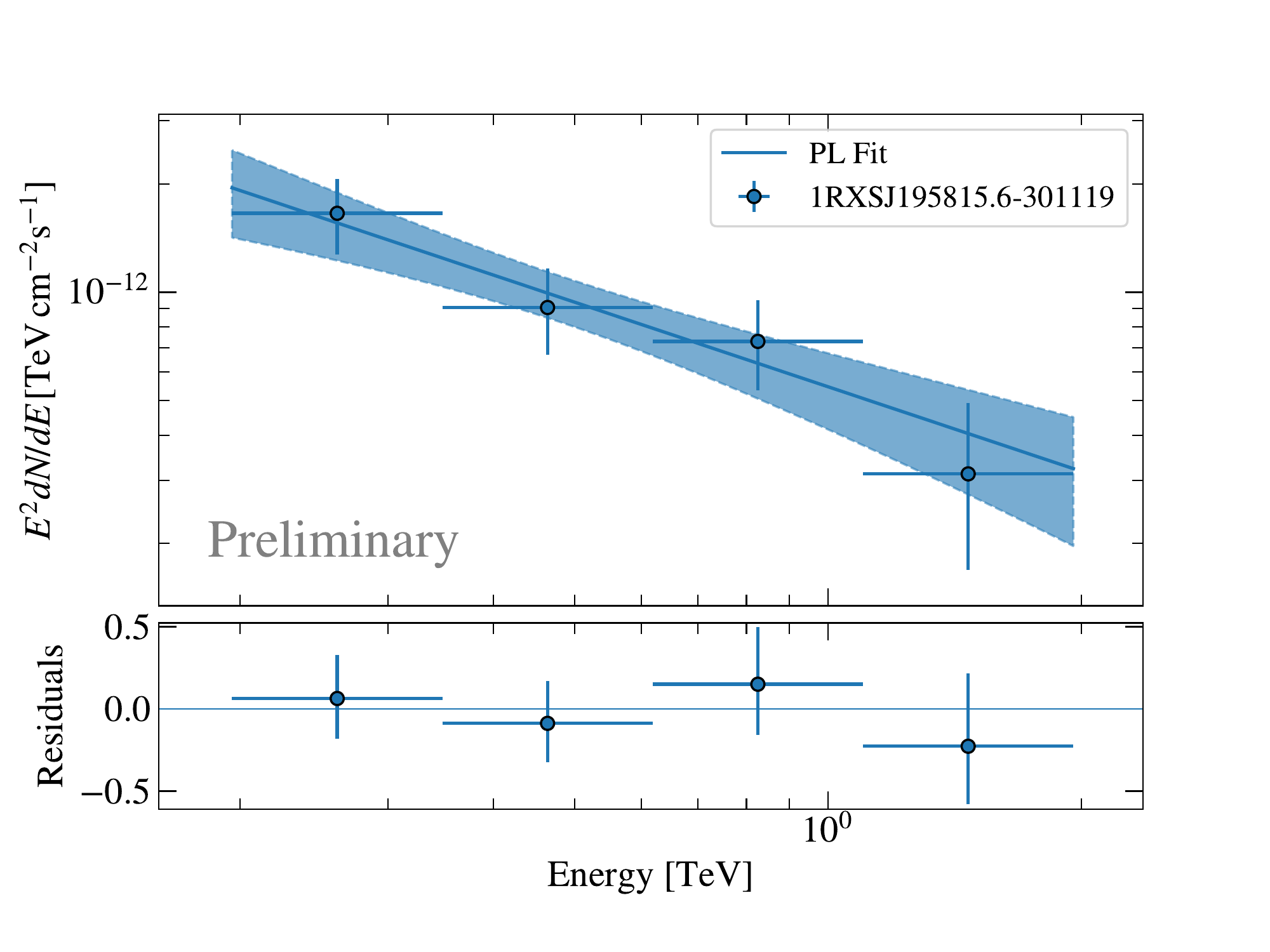}
    \caption{Very-high energy spectra measured with \hess of the two blazars considered in this work.
    The flux points are derived from forward folding and likelihood profiling in the restricted energy ranges using the flux point estimation method of \texttt{gammapy}. 
    }
    \label{fig:hess-spectra}
\end{figure}

In the energy range between 0.2 and 2\,TeV, the observed spectra are well described with simple power laws shown in Figure~\ref{fig:hess-spectra}, which are of the form $dN/dE = N (E/E_0)^{-\Gamma}$. The best-fit parameters are summarized in Table~\ref{tab:results}. Repeating the fit with power laws corrected for $\gamma$-ray absorption on the EBL (using the model in Ref.~\cite{2011MNRAS.410.2556D}) reveals hard intrinsic spectra emitted by these blazars with power-law indices compatible with $\Gamma_\mathrm{int} \approx 2$.

\section{\textit{Fermi}-LAT observations}

For the two sources, a \fermi-LAT analysis is performed over a time range of 11.5~years (August 8, 2008 - January 4, 2020) in order to extract average source spectra at energies above $100\,$MeV.
For the analysis, events are considered which pass the \texttt{P8R3\_SOURCE\_V2} selection criteria and have arrived within a region of interest (ROI) of $10^\circ\times10^\circ$ centered on the source position and with zenith angles $\leqslant 90^\circ$.
A spatial binning of $0.1^\circ$ per pixel and a logarithmic spectral binning of 8 bins per decade are chosen. 
The ROI is modelled by including all sources listed in the 4FGL catalog~\cite{2020ApJS..247...33A} up to $20^\circ$ away from the central source.
All the spectral parameters of sources within $6^\circ$ of the center are left free to vary, while only the spectral normalizations are free for sources  $6^\circ$-$10^\circ$ away from the source of interest.
Standard templates are used for the isotropic diffuse emission and Galactic diffuse emission.\footnote{\url{https://fermi.gsfc.nasa.gov/ssc/data/access/lat/BackgroundModels.html}}
The fit of the ROI is performed using \textsc{fermipy}~\cite{2017ICRC...35..824W} and the \emph{Fermi} Science tools.\footnote{See \url{http://fermipy.readthedocs.io/} and \url{https://fermi.gsfc.nasa.gov/ssc/data/analysis/documentation/}.}
The two sources are again described well with simple power laws and the best-fit parameters are reported in Table~\ref{tab:results}. Including EBL correction in the model only leads to a marginal change of the best-fit parameters as the fit is dominated by low-energy events for which the absorption is negligible. 

\section{X-ray observations}
The EHBLs have also been observed in X-rays with the XRT instrument on board the Neil Gehrels \swift observatory in photon counting mode. 
\rxsj has also been observed by \nustar. 
XRT operates in the energy range of 0.2-10\,keV \cite{2004ApJ...611.1005G} whereas \nustar observes X-rays in the range between 3 and 79\,keV \cite{2013ApJ...770..103H}.
The observation time of the two sources from XRT and \nustar are provided in Table~\ref{tab:results}. 

For the reduction and analysis of the observations, packages within HEASOFT 6.28 are used.
The spectral fitting is performed with XSPEC v12.11.1c.
Standard data reduction pipelines are used for both XRT and \nustar data using the pipelines  \textit{xrtpipeline} and \textit{nupipeline} (the latter being part of the \nustar Data Analysis Software package). 
After doing the analysis for each observation taken from XRT, the average spectra are obtained with the \swift-XRT data products generator~\cite{2009MNRAS.397.1177E}.\footnote{\url{https://www.swift.ac.uk/user_objects/}}  
The spectra from the two telescopes of \nustar, FMPA and FMPB, have been jointly fitted.

\begin{figure}[htb]
    \centering
    \includegraphics[width=.32\linewidth, angle = 90]{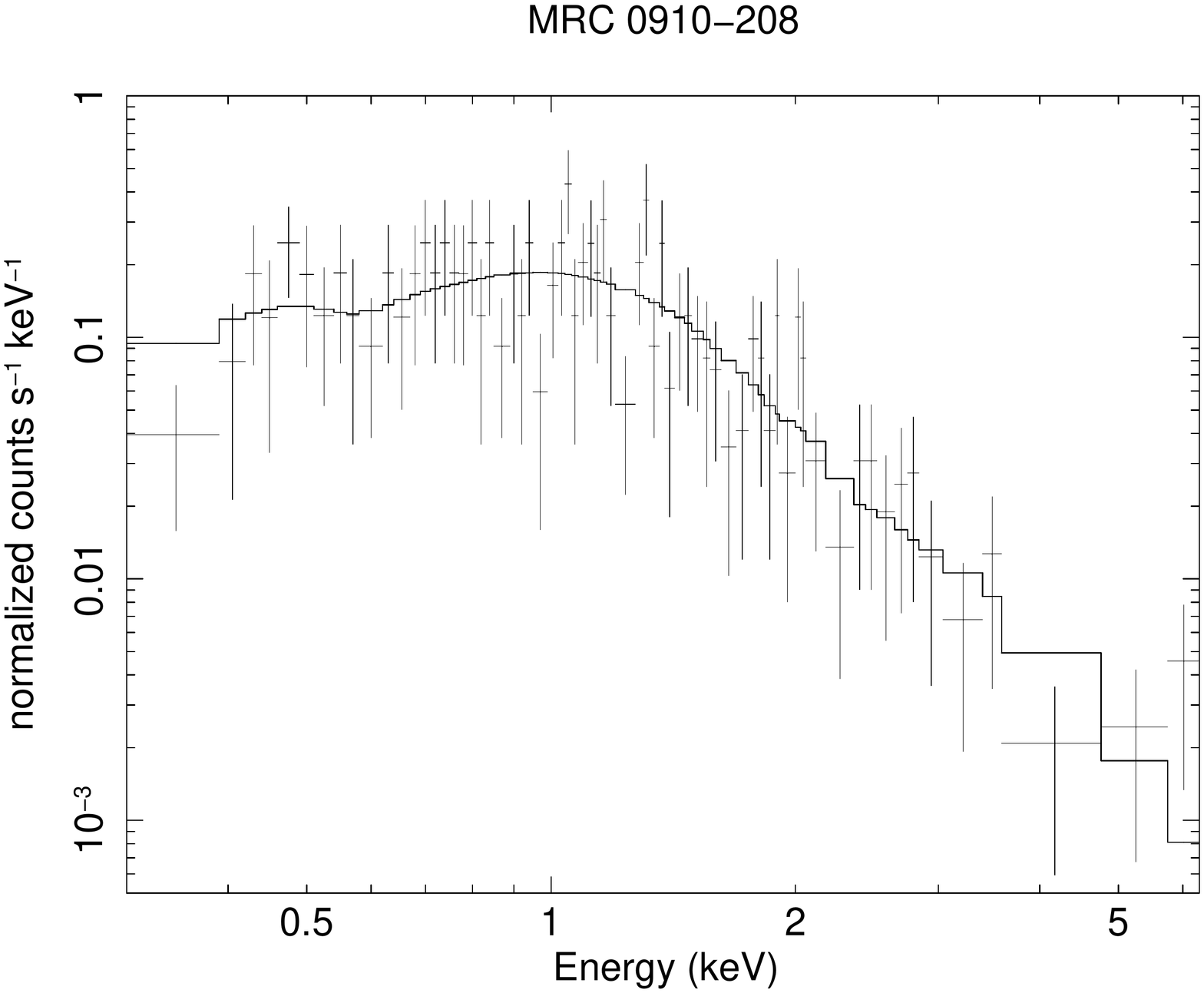}
    \includegraphics[width=.32\linewidth, angle = 90]{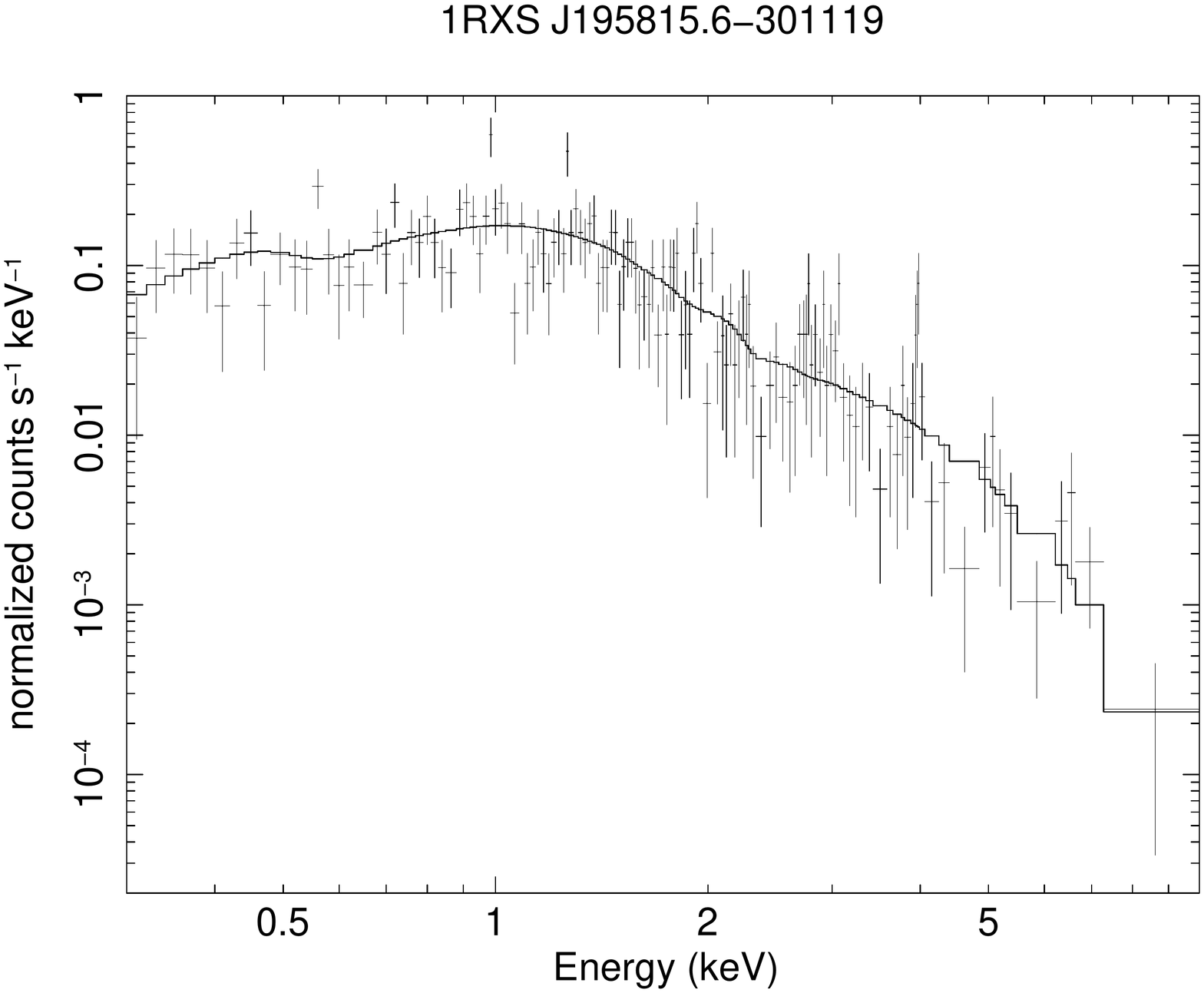}
    \includegraphics[width=.32\linewidth, angle = 90]{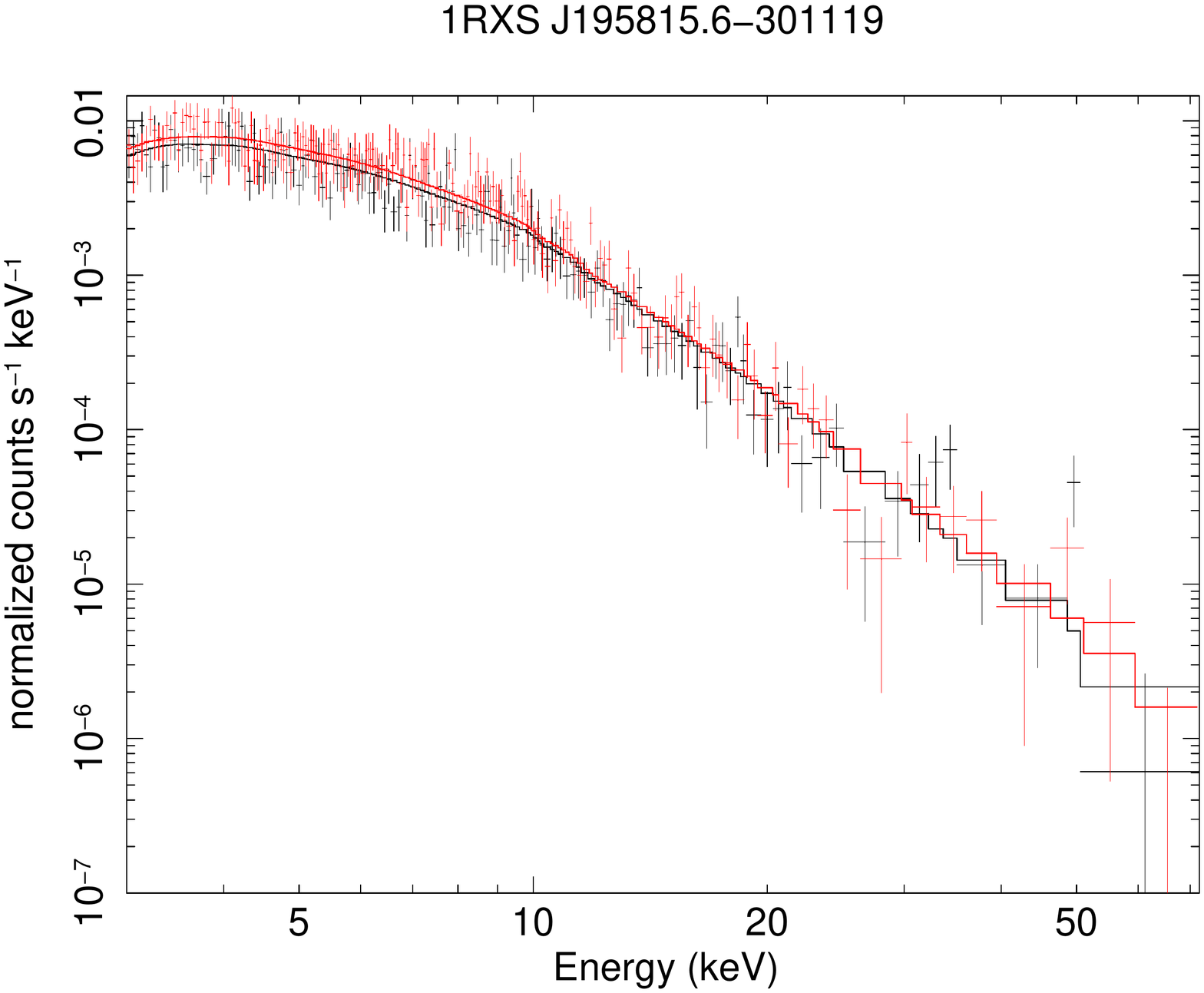}
    \caption{X-ray spectra of the two EHBLs. \textit{Top:} XRT spectra of \mrc and \rxsj. \textit{Bottom:} \nustar spectra of \rxsj. The \nustar spectra includes data from the FPMB telescope (shown in red) and the FPMA telescope (shown in black).}
    \label{fig:X-ray spectra}
\end{figure}

The spectral fitting has been done using a simple power-law model including Galactic absorption and corrected for redshift (\textit{zphabs} in XSPEC). Additionally, a log parabola model corrected with \textit{zphabs} has been tested in order to evaluate the synchotron peak energy.
The photon flux in the log parabola model is given by $dN/dE = N(E/E_0)^{-\alpha - \beta\ln(E/E_0)}$.
The hydrogen column density $N_\mathrm{H}$ for Galactic absorption has been fixed to the values of the LAB survey \cite{2005A&A...440..775K}.\footnote{\url{https://heasarc.gsfc.nasa.gov/cgi-bin/Tools/w3nh/w3nh.pl}}
Cash statistics (cstat in XSPEC) are used for the maximum likelihood estimation, which is suitable  for sources with low photon counts \cite{1979ApJ...228..939C}.
The spectral fitting parameters are given in Table~\ref{tab:results}. 
For both sources, a slight preference for the log-parabola (LP) model over the power-law (PL) model is found from the difference in the Cash statistics, $\Delta C = C_\mathrm{PL} - C_\mathrm{LP}$. 
The LP model is preferred with a statistical significance of $2.3\,\sigma$ and $2.8\,\sigma$ for \mrc and \rxsj, respectively. 
For \nustar, only the power-law parameters are provided, as the log parabola fit is not significantly preferred. 
The best-fit spectra are also shown in Figure~\ref{fig:X-ray spectra}.
The peak of the synchrotron emission is estimated from the maximum of the best-fit log parabola. The peak energy is given by $E_\mathrm{peak} = E_0 \exp(\alpha / (2\beta))$ and the uncertainty is derived with Gaussian error propagation. 

\section{Modeling of the multi-wavelength spectral energy distribution}
\label{sec:ssc}

\begin{figure}[htb]
    \centering
    \includegraphics[width=.49\linewidth]{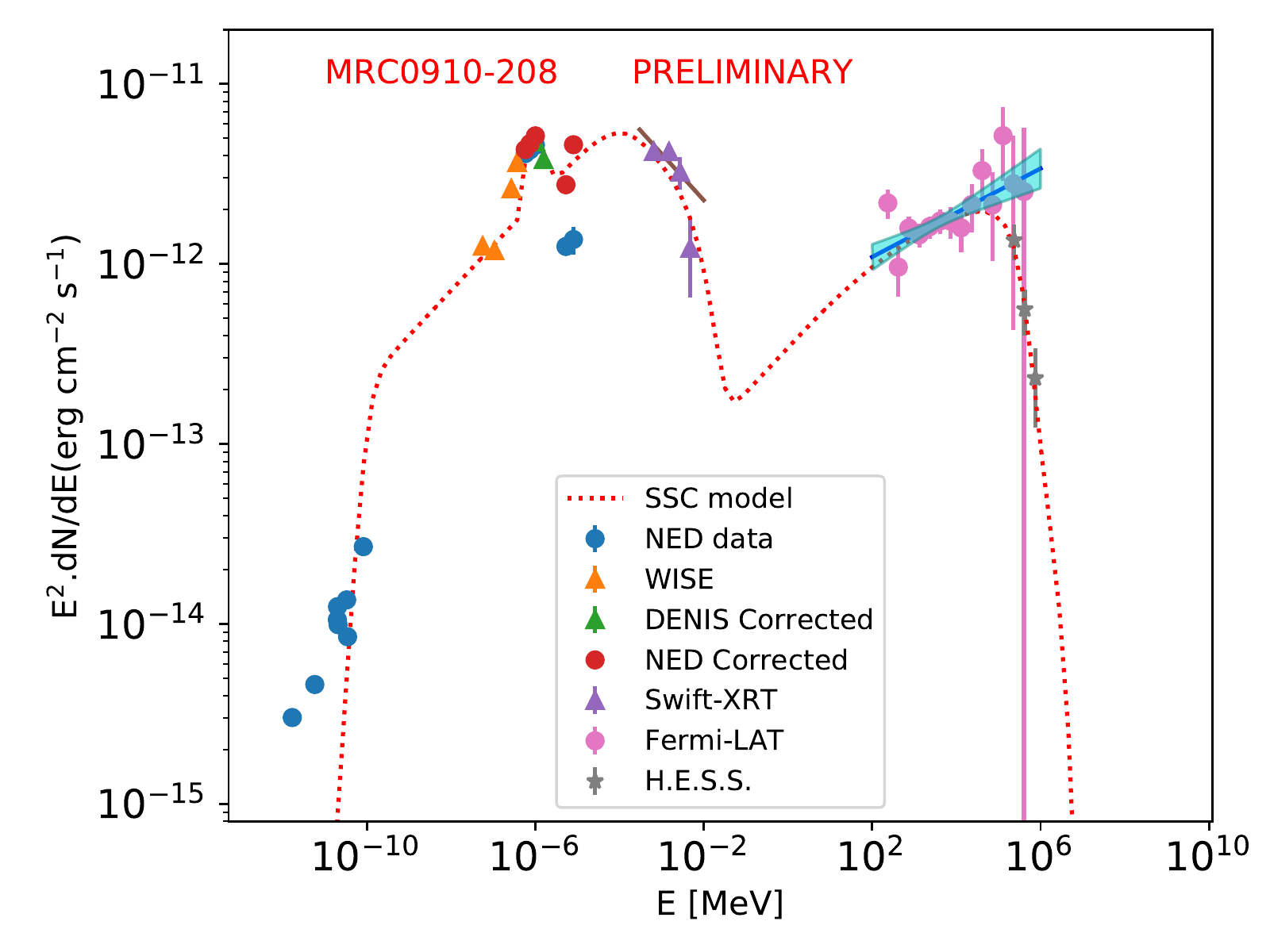}
    \includegraphics[width=.49\linewidth]{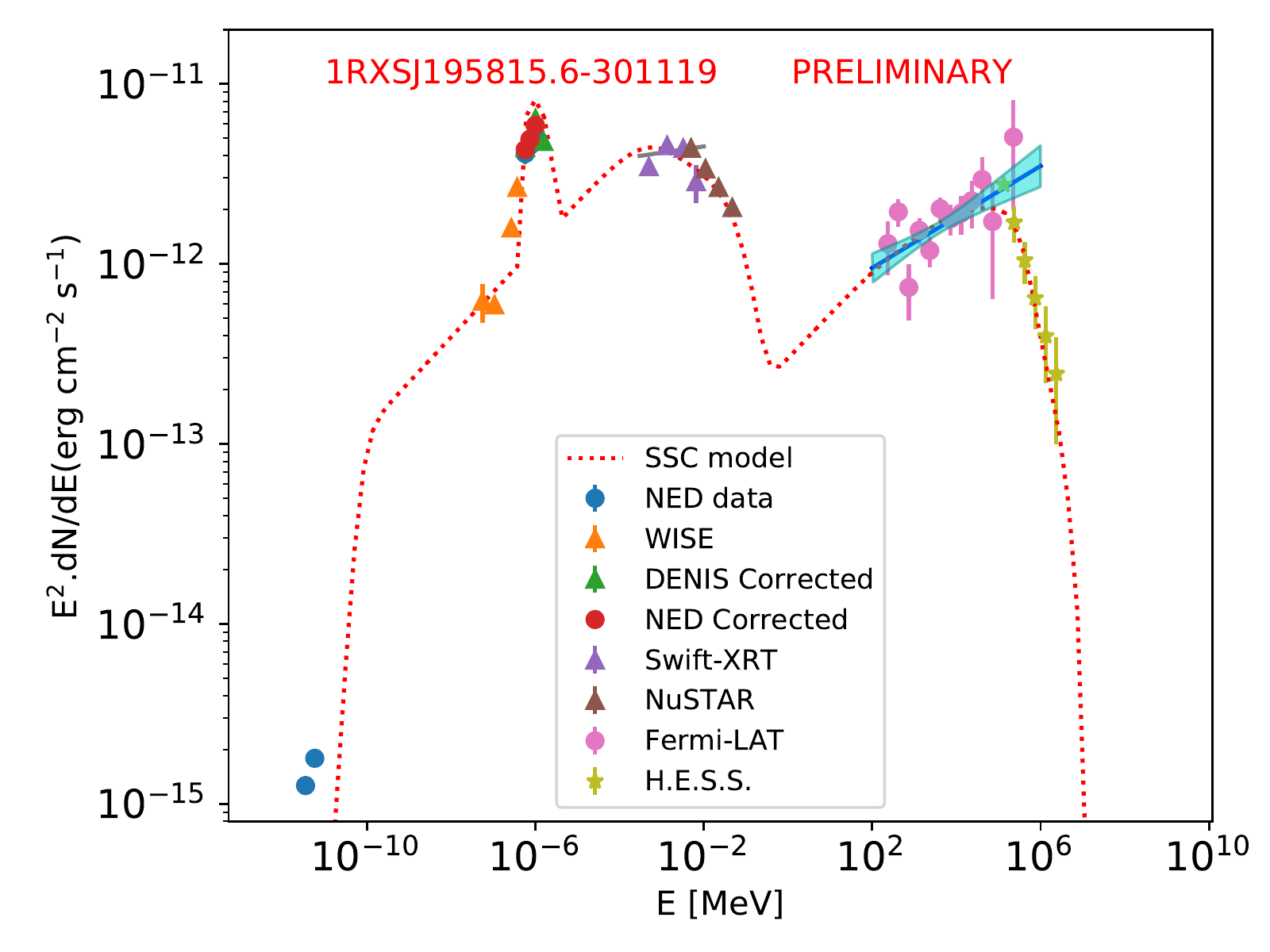}
    \caption{SEDs of the EHBLs presented in this work.
    Optical data were extracted from NED, WISE, GALEX and DENIS. X-ray, \fermi and H.E.S.S. data are presented in this work. The dashed line is the SSC calculation, which includes absorption on the EBL. 
    }
    \label{fig:sed}
\end{figure}

The multi-wavelength SEDs of the sources are shown in Fig.~\ref{fig:sed}.
Archival data from the NASA/IPAC Extragalactic Database (NED)\footnote{\url{http://ned.ipac.caltech.edu/}} together with data from WISE \cite{2013wise.rept....1C}, DENIS~\cite{1997Msngr..87...27E}, and GALEX~\cite{2007ApJS..173..682M} are used along with the X-ray and $\gamma$-ray data presented in this work.
A one-zone SSC model is used in order to reproduce the SEDs.
The calculations are performed with the \texttt{agnpy} python package.\footnote{\url{https://agnpy.readthedocs.io/en/latest/}} 
The SSC model assumes a spherical emission zone of size $R$, which is filled with a uniform magnetic field of strength $B$ and moves down the jet with a bulk Lorentz factor $\Gamma_\mathrm{L}$. 
For simplicity, we assume the bulk Lorentz factor is equal to the Doppler factor.
The lepton distribution $dN_e/d\gamma$ is a function of the electron and positron Lorentz factor $\gamma$ and is  assumed to follow a broken power law ( with index $\alpha_1$ and $\alpha_2$) with a break at $\gamma_b$. It is defined between $\gamma_\mathrm{min}$ and $\gamma_\mathrm{max}$.

For the two blazars, a thermal component in the SED is visible and a giant-elliptical template representing the radiation from the AGN's host galaxy (extracted from the ASDC SED builder\footnote{\url{https://tools.ssdc.asi.it/SED/}}) is added to the model.
The parameters of the model are reported in the bottom part of Table~\ref{tab:results}.
For all sources, the SSC model is far from equipartition and is dominated by the particle energies. In the modelling, we ensure that the peak of the synchrotron emission is in agreement with the peak of the multi-wavelength SED (obtained with a fit by a polynomial function in log-log).

\section{Conclusions}

\hess detections at very-high-energy $\gamma$-ray energies of two EHBLs, \mrc and \rxsj, have been presented.
Using archival \swift-XRT observations, the EHBL nature of these sources could be confirmed by confidently placing the synchrotron peak energy above 1\,keV or a frequency beyond $\approx 2.4\times10^{17}\,$Hz. 
Both sources also exhibit hard intrinsic $\gamma$-ray spectra that can be well described with power laws with indices compatible with $\Gamma_\mathrm{int} \approx 2$. 
Hard spectra are also measured with \fermi LAT with $\Gamma < 2$. 
We have also performed a simple SSC modeling of the sources.
As mentioned in the introduction, Klein-Nishina suppression can become important at energies above 1\,TeV.
However, the current data sets are rather well described with the SSC model,
which also reveals that both sources can be described with similar parameters for the underlying particle distribution. 
Further data taking might be required to probe the emission at energies beyond 1\,TeV for these sources.
A dedicated modeling also considering hadronic interactions as done in, e.g., Ref.~\cite{2015MNRAS.448..910C} together with an analysis of CT5 data will be presented elsewhere.
The detections reported here further increase the the growing sample of EHBLs~\cite[see, e.g.,][for recent MAGIC detections]{2020ApJS..247...16A}.
This will facilitate future constraints on the EBL in the IR as well as searches for pair cascades produced in the intergalactic medium. 

\textit{Acknowledgements.}
M.  M.  acknowledges  support from the European Research Council (ERC) under the European Union’s Horizon 2020 research and innovation program Grant agreement No. 843800 (GammaRayCascades) and No. 948689 (AxionDM).
The \textit{Fermi}-LAT Collaboration acknowledges support for LAT development, operation and data analysis from NASA and DOE (United States), CEA/Irfu and IN2P3/CNRS (France), ASI and INFN (Italy), MEXT, KEK, and JAXA (Japan), and the K.A.~Wallenberg Foundation, the Swedish Research Council and the National Space Board (Sweden). Science analysis support in the operations phase from INAF (Italy) and CNES (France) is also gratefully acknowledged. This work performed in part under DOE Contract DE-AC02-76SF00515.

This work made use of data supplied by the UK Swift Science Data Centre at the
University of Leicester.

\bibliographystyle{jhep}
\bibliography{mainbib}

\clearpage
\section*{Full Authors List: H.E.S.S.}

\scriptsize
\noindent
H.~Abdalla$^{1}$, 
F.~Aharonian$^{2,3,4}$, 
F.~Ait~Benkhali$^{3}$, 
E.O.~Ang\"uner$^{5}$, 
C.~Arcaro$^{6}$, 
C.~Armand$^{7}$, 
T.~Armstrong$^{8}$, 
H.~Ashkar$^{9}$, 
M.~Backes$^{1,6}$, 
V.~Baghmanyan$^{10}$
% ...

\medskip

\noindent
$^{1}$University of Namibia, Department of Physics, Private Bag 13301, Windhoek 10005, Namibia\\
$^{2}$Dublin Institute for Advanced Studies, 31 Fitzwilliam Place, Dublin 2, Ireland\\
$^{3}$Max-Planck-Institut f\"ur Kernphysik, P.O. Box 103980, D 69029 Heidelberg, Germany\\
$^{4}$High Energy Astrophysics Laboratory, RAU,  123 Hovsep Emin St  Yerevan 0051, Armenia\\
$^{5}$Aix Marseille Universit\'e, CNRS/IN2P3, CPPM, Marseille, France\\
$^{6}$Centre for Space Research, North-West University, Potchefstroom 2520, South Africa\\
$^{7}$Laboratoire d'Annecy de Physique des Particules, Univ. Grenoble Alpes, Univ. Savoie Mont Blanc, CNRS, LAPP, 74000 Annecy, France\\
$^{8}$University of Oxford, Department of Physics, Denys Wilkinson Building, Keble Road, Oxford OX1 3RH, UK\\
$^{9}$IRFU, CEA, Universit\'e Paris-Saclay, F-91191 Gif-sur-Yvette, France\\
$^{10}$Instytut Fizyki J\c{a}drowej PAN, ul. Radzikowskiego 152, 31-342 Krak{\'o}w, Poland\\
% ...

%
%
%

\end{document}